# Brief Announcement: Decentralized Construction of Multicast Trees Embedded into P2P Overlay Networks based on Virtual Geometric Coordinates

Mugurel Ionuţ Andreica, Andrei Drăguş, Ana-Delia Sâmbotin, Nicolae Ţăpuş
Computer Science Department, Politehnica University of Bucharest
Splaiul Independenţei 313, sector 6, Bucharest
{mugurel.andreica, ntapus}@cs.pub.ro, andreidragus@yahoo.com, delia.sambotin@gmail.com

## ABSTRACT
In this paper we consider the problem of efficiently constructing in a fully distributed manner multicast trees which are embedded into P2P overlays using virtual geometric node coordinates. We consider two objectives: to minimize the number of messages required for constructing a multicast tree by using the geometric properties of the P2P overlay, and to construct stable multicast trees when the lifetime durations of the peers are known.

## Categories and Subject Descriptors
C.2.1 [**Computer-Communication Networks**]: Network Architecture and Design – *distributed networks, network topology;* C.2.4 [**Computer-Communication Networks**]: Distributed Systems – *distributed applications*

## 1. INTRODUCTION
Many of the existing multicast tree solutions are very sensitive to node departures, send many messages for constructing the tree or are not fully decentralized. In this paper we investigate the problem of constructing in a fully distributed manner multicast trees embedded into P2P overlays in which the nodes have virtual geometric coordinates assigned to them. We present a multicast tree construction algorithm which makes use of these coordinates in order to send a minimum number of messages. We also consider the situation in which every participating node $i$ knows the time moment $T(i)$ when it will leave the system and we use this information for constructing stable multicast trees. This information may be available in: (1) *Cloud computing* (the nodes are applications running on virtual machines which are leased for fixed periods of time) and (2) *Wireless sensor networks* (the sensors know the remaining lifetime of their battery).

Every peer $i$ is assigned a self-generated identifier $(x(i,1), …, x(i,D))$, representing the coordinates of a point in a D-dimensional space (all the coordinates are between $0$ and *VMAX*). W.l.o.g., we assume that all the coordinates in the same dimension are distinct. The peers are connected into a P2P overlay which makes use of the peers' identifiers [1]. When a peer $P$ joins the system, it must know the identifiers and the network addresses (public IP and port) of one or more peers which are already within the system. These peers are the initial neighbours of P. Then, periodically, every peer broadcasts its existence (i.e. its identifier and network address) a fixed number $BR≥2$ of hops away from it within the P2P overlay. Let $I(P)$ be the set of peers from which the peer $P$ received existence announcements during the previous *Tmax* seconds (where *Tmax* is larger than the gossiping period). Then, using a *neighbour selection method*, $P$ selects its new overlay neighbours from $I(P)$. This method must ensure that, as long as no new peers enter the system and no old peers leave the system, the overlay topology converges to an equilibrium. If the peers enter or leave the system one at a time and the topology converges between two such events, then the equilibrium topology after every event should be the same (or "close to") the one obtained when every peer $P$ knows all the other peers in the system (i.e. when $I(P)$ contains all the peers except $P$). The *Hyperplanes neighbour selection method* [1] uses a set of $H$ hyper-planes, all of which contain the origin of the coordinate system. A peer $P$ conceptually translates the coordinates of the peers $Q$ from $(I(P) \cup \{P\})$ such that the coordinates of $P$ become the origin of the coordinate system. The $H$ hyper-planes divide the space into multiple regions. $P$ selects the $K$ closest peers to the conceptual origin (i.e. $P$) from each region as its neighbours (using a distance function). Three instances of this method are: 1) The set of hyper-planes consists of $d$ orthogonal hyper-planes – the $i^{th}$ ($1≤i≤D$) hyper-plane is: $x(i)=0$ (the *Orthogonal Hyperplanes* method); 2) The hyper-planes are: $a(1)·x(1)+…+a(D)·x(D)=0$, where the coefficients $a(i)$ may be $-1$, $0$, or $+1$ [2] ; 3) $H=0$ (there is only one region; the $K$ closest peers from $I(P)$ are selected).

## 2. MULTICAST TREES BASED ON SPACE PARTITIONING TECHNIQUES
In this section we present a generic method for constructing multicast trees embedded in a geometric P2P overlay, using space partitioning techniques. The proposed algorithm uses the concept of *responsibility zone* $Z(P)$ of a peer $P$. $Z(P)$ represents a zone from the D-dimensional space, with the property that the peer $P$ will be responsible for delivering the multicast data (directly or indirectly) to all the peers in this zone. Let $A$ be the peer which initiates the multicast tree construction. $Z(A)$ will be the entire virtual coordinate space. The algorithm works as follows. Let's assume that a multicast tree construction request message arrived at a peer $P$ (we will assume that $A$ also receives such a message, implicitly). The message contains the description of $Z(P)$. $P$ will select a subset of its neighbours from the overlay topology which are located within Z(P), which will become its tree neighbours. For each selected neighbour Q, P will compute its responsibility zone Z(Q). Q will be contained within Z(Q) and no other selected neighbour of P will belong to Z(Q). The zones Z(Q) of the selected neighbours Q are disjoint. Their union must contain all the peers from Z(P) which have not received the message, yet, and must exclude P (i.e. the identifier of P should be outside of their union), in order to send the message to every peer in Z(P) and to avoid receiving duplicate messages. Then, P will forward the request message to the selected neighbours Q (the message forwarded to Q contains the description of Z(Q)). The algorithm sends N-1 messages, where N is the total number of peers.

In order to test our algorithm, we developed a multi-threaded simulation framework in Python. The implemented neighbour selection method was the following. Each peer P selects as neighbours all the peers Q from the set I(P), such that the axes-aligned hyper-rectangle having as corners the identifiers of P and Q contains no other identifier of a peer from I(P) inside (i.e. the hyper-rectangle whose side in each dimension $i$ is $[min\{x(P,i), x(Q,i)\}, max\{x(P,i), x(Q,i)\}]$ ($1 \leq i \leq D$) contains no other peer from I(P)). The responsibility zone of each peer P was always the (strict) interior of an axes-aligned hyper-rectangle. When P receives the message, it classifies all of its neighbours as in the *Orthogonal Hyperplanes* method, according to the region to which they belong. Then, within each region, the neighbours are sorted in increasing order of the $L_1$ distance to P. From each region, the peer Q with the median distance to P is selected. Z(Q) is computed as the intersection of Z(P) with the hyper-rectangle HR corresponding to the region to which Q belongs, relative to P. The side of HR in dimension $i$ is: *if $(x(Q,i)<x(P,i))$ then $(-\infty, x(P,i))$ else $(x(P,i), +\infty)$ ($1 \leq i \leq D$)*.

We ran multiple tests with *N=1000* peers, for *D=2..5*. The coordinates of each peer were randomly generated and the peers were inserted one by one in the overlay (the overlay was allowed to converge after every insertion). For each test we computed the maximum and average degree of a peer within the obtained P2P topology. Then, we initiated the multicast tree construction from each of the N peers and, in each case, we computed the longest path from the initiating peer (the root) to a leaf in the tree. We also computed the average length of the longest path over all the N multicast sessions. Fig. 1 (a-b) presents the results. The best trade-off between overlay node degree and tree path lengths occurs for *D=2*, when both the maximum and average degree of a peer in the overlay topology seem to be proportional to *log(N)* (see Fig. 1 - c, which presents simulation results for *D=2*, random peer coordinates and various numbers of peers). The maximum tree degree of a peer was bounded by $2^D$, as expected.

## 3. MULTICAST TREES WITH IMPROVED STABILITY PROPERTIES

In this section we make the following supplementary assumption. Every peer P also knows the time moment T(P) when it will leave the P2P topology. Then, we set x(P,1)=T(P) (i.e. the $1^{st}$ coordinate of P will be T(P)). We will make the simplifying assumption that the clocks of all the peers are synchronized and that all the values T(*) are distinct (this can be easily achieved, by breaking ties based on other peer-specific properties). The multicast tree construction algorithm works as follows. Among all of its overlay neighbours, every peer P will (periodically) select a peer Q with T(Q)>T(P) as its *preferred tree neighbour*. If no such peer Q exists, then P selects no preferred tree neighbour. Among all the neighbours Q of P with T(Q)>T(P), the peer P may select any of them as the preferred tree neighbour (secondary selection criteria may be used). The algorithm constructs a tree in which peers P with lower values of T(P) are located towards the edge of the tree (closer to the leaves). Thus, when a peer P leaves the tree, P will be a leaf in the tree and the tree will not be disconnected by P's departure. We tested our algorithm using the simulation framework mentioned previously. The T(*) values of the peers and their coordinates were randomly generated. We used the *Orthogonal Hyperplanes* neighbour selection method for constructing the P2P overlay. The preferred tree neighbour of a peer P was selected to be the overlay neighbour Q with the largest value T(Q), if T(Q)>T(P). We simulated the construction of the P2P topology (the peers were inserted as before) and, after reaching the equilibrium, we constructed the multicast tree. We used *N=1000* peers and multiple values of D (*2..10*) and K (*1..50*). In each case, the preferred neighbour links indeed formed a tree. We rooted the tree at the peer P with the largest value T(P), thus defining parent-son relationships, and then we checked that T(A)>T(B) for every two peers A and B such that A is the parent of B (i.e. that the values T(*) decrease as we advance towards the leaves of the tree). The mentioned conditions were always met. Fig. 1 (d-e) presents the results. For small values of K, both the maximum degree and the tree diameter are quite small.

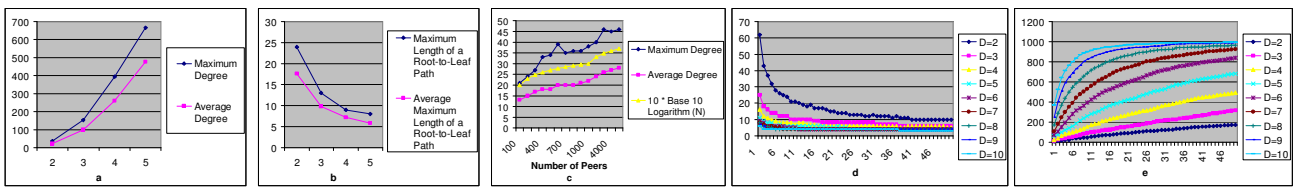

**Figure 1. a-Maximum and average topology degree of a peer for D=2..5 (N=1000). b-Maximum length and average maximum length of a root-to-leaf path in the tree for D=2..5 (N=1000). c-Maximum and average topology degree of a peer for D=2 and N=100..5000. d-Variation of the multicast tree diameter with K (1..50) for multiple values of D (2..10). e-Variation with K (1..50) of the maximum degree in the multicast tree of a peer, for multiple values of D (2..10).**